\documentclass[%
 reprint,
 amsmath,amssymb,
 aps,
]{revtex4-1}

\usepackage{graphicx}% Include figure files
\usepackage{dcolumn}% Align table columns on decimal point
\usepackage{bm}% bold math
\usepackage{CJK}
\usepackage {soul}
\usepackage{graphicx}
\usepackage{subfigure}
\usepackage{caption}
\usepackage{subfig,graphicx}

\begin{document}
%\pagewiselinenumbers
%\switchlinenumbers

\preprint{APS/123-QED}

\title{Microscopic study of deformation and orientation effects in heavy-ion reactions above Coulomb barrier using the Boltzmann-Uehling-Uhlenbeck model}% Force line breaks with \\
%\thanks{A footnote to the article title}%
\author{Yujie Feng\textsuperscript{1}}\thanks{the authors contributed equally}

% \altaffiliation[Also at ]{Physics Department, XYZ University.}%Lines break automatically or can be forced with \\

\author {Huizi Liu\textsuperscript{1}}\thanks{the authors contributed equally}
\author {Yingge Huang\textsuperscript{1}}%
\author {Fuchang Gu\textsuperscript{1}}%
\author {Erxi Xiao\textsuperscript{1}}
\author {Xin Lei\textsuperscript{1}}%
\author {Hui Wang\textsuperscript{1}}
\author {Jiali Huang\textsuperscript{1}}
\author {Long Zhu\textsuperscript{1,3}}
\author {Jun Su\textsuperscript{1,2,3}}
\email{sujun3@mail.sysu.edu.cn}
\affiliation{%
 Sino-French Institute of Nuclear Engineering and Technology, Sun Yat-sen University, Zhuhai 519082, China
}
\affiliation{China Nuclear Data Center, China Institute of Atomic Energy, Beijing 102413, China}
\affiliation{Guangxi Key Laboratory of Nuclear Physics and Nuclear Technology,
Guangxi Normal University, Guilin 541004, China}
%

%\collaboration{MUSO Collaboration}%\noaffiliation

%\author{Charlie Author}
% \homepage{http://www.Second.institution.edu/~Charlie.Author}
%\affiliation{
Second institution and/or address\\
% This line break forced% with \\
%}%
%\affiliation{
% Third institution, the second for Charlie Author
%}%
%\author{Delta Author}
%\affiliation{%
% Authors' institution and/or address\\
% This line break forced with \textbackslash\textbackslash
%}%

%\collaboration{CLEO Collaboration}%\noaffiliation

\date{\today}% It is always \today, today,
             %  but any date may be explicitly specified

\begin{abstract}
\begin{description}
\item[Background] The understanding of the impact of initial deformation and collision orientation on quasi-fission and fusion-fission reactions remains incomplete.

\item[Purpose] This article aims to explore how the orientation of deformed nuclei influences quasi-fission and fusion-fission around 1.2 $V_B$, employing a microdynamical method in systems with diverse shapes, namely $^{24}$Mg + $^{178}$Hf, $^{34}$S + $^{168}$Er, and $^{48}$Ti + $^{154}$Sm.

\item[Method] Utilizing the Boltzmann-Uehling-Uhlenbeck model, this study investigates quasi-fission and fusion-fission reactions. The model elucidates micro-dynamic processes and microscopic observables through the definition of the window and event-by-event simulations.

\item[Results] The findings reveal that the orientation of deformed nuclei significantly influences the nucleus-nucleus interaction potential, thereby impacting the competition between quasi-fission and fusion-fission. Particularly, the orientation of the deformed target nucleus emerges as the primary factor affecting this competition. Notably, a higher proportion of fusion-fission events is observed when the target nucleus is in the belly orientation compared to the tip. The study also observes that the configuration of the dinuclear system contributes to fluctuations and dissipation. Collisions with different orientations result in distinct dinuclear system configurations, with belly-oriented collisions leading to larger fluctuations between events, while tip-oriented collisions exhibit smaller fluctuations.

\item[Conclusions] Considering diverse orientations of nuclei with distinct initial deformations, this study concludes that the orientation of the target nucleus is the key factor influencing quasi-fission and fusion-fission reactions around 1.2 $V_B$. 
%\item[Usage] Secondary publications and information retrieval purposes.
%\item[PACS numbers]
%May be entered using the \verb+\pacs{#1}+ command.
%\item[Structure] You may use the \texttt{description} environment to structure your abstract; use the optional argument of the \verb+\item+ command to give the category of each item.
\end{description}
\end{abstract}

\pacs{Valid PACS appear here}% PACS, the Physics and Astronomy
                             % Classification Scheme.
%\keywords{Suggested keywords}%Use showkeys class option if keyword
                              %display desired
\maketitle
\section{\label{int}Introduction}

Quasi-fission (QF) is a pivotal process in heavy-ion reactions, characterized by the formation of asymmetric fragments \cite{giuliani2019colloquium}.
Its significance arises from its competition with fusion, a phenomenon critical to the understanding of superheavy element synthesis. 
The propensity of QF to lead to asymmetric fragment formation poses a challenge to the desired formation of compound nuclei, thus garnering considerable attention from researchers in the past few decades \cite{simenel2021comparison}.
Consequently, a thorough investigation into the competition between QF and fusion is indispensable for optimizing reaction conditions and enhancing the probability of synthesizing superheavy elements \cite{hinde2021experimental}.
Furthermore, the non-equilibrium nature of QF reactions offers a unique advantage. The observables of the final state retain crucial information about the entrance channel conditions, providing valuable insights into the effects of the initial nuclear configurations \cite{hammerton2019entrance}. 
This characteristic makes QF an invaluable tool for studying the intricate interplay of factors influencing heavy-ion reactions, contributing to our understanding of the dynamics involved in the synthesis of superheavy elements.

The influence of entrance channel effects on reaction dynamics near the Coulomb barrier has captivated the attention of both experimental and theoretical researchers. 
In the experimental realm, extensive investigations have been undertaken to discern entrance channel conditions favoring QF. 
Noteworthy factors under scrutiny include mass asymmetry \cite{berriman2001unexpected}, fissionability of compound nuclei \cite{lin2012systematic}, reaction energy \cite{back1985complete}, magicity \cite{simenel2012influence}, and neutron richness \cite{adamian2000isotopic}.
In the theoretical domain, the macroscopic dinuclear system model has been pivotal in revealing the impact of orientation effects on the distribution of the nucleus-nucleus interaction potential \cite{li2005deformation}. 
Furthermore, dynamic simulations using microscopic models such as time-dependent Hartree-Fock (TDHF) and stochastic mean-field (SMF) have provided valuable insights into reaction dynamics under various initial conditions \cite{sekizawa2017enhanced, sekizawa2016time, zheng2018connecting, ayik2019heavy, ayik2019quantal}. 
The microscopic ImQMD model, in particular, has contributed significantly to the exploration of the effects of initial deformation and orientation on QF probability, as evidenced by studies such as \cite{zhao2013production,wang2015fusion}.
These theoretical frameworks not only shed light on the intricate interplay of entrance channel parameters but also offer a nuanced understanding of the underlying mechanisms governing QF. 

In recent decades, researchers have delved into the intricate interplay of initial nuclear deformation and orientation in QF and fusion-fission (FF) reactions. 
The seminal work in 1995 by Hinde et al. demonstrated a correlation between nuclear deformation, orientation, and the competition of QF and FF \cite{hinde1995fusion}. 
Subsequent investigations by S. Mitsuoka et al. in 2002 brought attention to the role of an 'extra push' in tip collisions, where additional kinetic energy was required for fusion. 
This extra energy was associated with the elongated contact configuration and the fission saddle-point configuration. 
In contrast, belly collisions required less of this 'extra push' for fusion reactions \cite{mitsuoka2002effects}. 
A pivotal experiment conducted by Knyazheva et al. in 2007 underscored the crucial role of target nucleus deformation throughout the entire reaction \cite{knyazheva2007quasifission}.
Comparing experimental data on QF with theoretical models has been instrumental in testing the predictive capabilities of these models and probing the dynamical processes involved. 
The TDHF calculations by Simenel et al. in 2012 revealed that tip collisions predominantly result in the QF component \cite{simenel2012effects}.
Subsequent the TDHF calculations by Seikizawa et al. indicated that collisions with different orientations influence neck nucleon transfer in the dinuclear system, thereby affecting reactions near the Coulomb barrier \cite{sekizawa2017enhanced}. 
Recent TDHF calculations have extended this understanding, highlighting the impact of the initial orientation of a deformed nucleus on intermediate configurations and the dynamical trajectory of QF reactions \cite{mcglynn2023time}.

While significant progress has been made, the specific dynamical factors governing the competition between QF and FF are not yet fully elucidated. Further investigation is needed to uncover the nuanced interactions influencing this competition and refine our understanding of the underlying mechanisms in heavy-ion reactions.
In our prior study, we introduced and applied the Boltzmann-Uehling-Uhlenbeck (BUU) model to examine reactions involving nuclei near the Coulomb barrier. 
Our findings demonstrated substantial agreement with experimental data, as detailed in \cite{feng2023contributions}. 
Building on this work, our current research extends the inquiry to explore the nuanced influence of orientation and deformation in the dynamical mechanisms governing the competition between QF and FF reactions around 1.2 times the Coulomb barrier energy ($V_B$).
The paper is organized as follows. In Sec. II, we briefly describe the theoretical method. In Sec. III, we present both results and discussions. Finally, the conclusions and overviews are given in Sec. IV.

\section{\label{model}Theoretical framework}

The development of successive approximations for the original global N-body problem is based on the concept of the BBGKY hierarchy \cite{durand2000nuclear}. By selectively excluding specific terms in the first equation of the hierarchy using the Hartree-Fock approximation and ignoring the two body correlation, we can derive the TDHF equation, which is a time-dependent formulation of the Hartree-Fock method \cite{dirac1930note}:

\begin{equation}
  i\frac{\partial \hat{\rho}}{\partial t} = \left[ \hat{h}\left[\hat{\rho}\right], \hat{\rho} \right],
  \label{THDF}
\end{equation}
where $\hat{\rho}$ represents the one-body density matrix, while $\hat{h}$ corresponds to the mean-field Hamiltonian. By making the assumption of molecular chaos, it is possible to solve the second equation of the hierarchy and subsequently express the first equation of the hierarchy as the quantum Boltzmann equation \cite{reinhardt1985non, reinhard1986dissipative},
\begin{equation}
  i\frac{\partial \hat{\rho}}{\partial t} = \left[ \hat{h}\left[\hat{\rho}\right], \hat{\rho} \right] + \hat{I}\left[\hat{\rho}\right],
  \label{QBE}
\end{equation}
where $\hat{I}$ represents the collision term and comes from the solution of the second equation of the hierarchy.

In the quantum framework, dealing with the collision term for finite fermion systems can be extremely challenging \cite{reinhard2015quantum, dinh2022quantum}. Therefore, in the field of heavy-ion physics, the Wigner representation is widely utilized, which resulting in the classical Boltzmann equation.
\begin{equation}
  \left( \frac{\partial}{\partial t} + \frac{\mathbf{p}}{m} \cdot \nabla_{r} - \nabla_{r}U(f)\cdot \nabla_{p} \right)f(\mathbf{r}, \mathbf{p}, t) = I(f).
  \label{BUU}
\end{equation}
here, $f$ represents the single-particle density in the phase space, encompassing the degrees of freedom of coordinate $\mathbf{r}$ and momentum $\mathbf{p}$. The left-hand side of the equation describes the evolution of the single-particle density, which is influenced by the mean-field potential $U$. This potential incorporates both the nuclear and Coulomb interactions.
The term on the right-hand side, i.e. $I(f)$, denotes the collision term of the Uehling-Uhlenbeck form, in which the Pauli principle is considered.
%   \begin{equation}
%   \begin{aligned}
%      I(f_{1}) = \int &\left[f_{3}f_{4}(1-f_{1})(1-f_{2})-f_{1}f_{2}(1-f_{3})(1-f_{4})\right]\\ & W(12;34) d\mathbf{p}_{2} d\mathbf{p}_{3} d\mathbf{p}_{4} ,
%      \label{K}
%   \end{aligned}
%   \end{equation}
% where $f_{i}$ = $f(\mathbf{r}, \mathbf{p}_{i}, t)$ is the phase space density of the $i$th particle and $W(12;34)$ is the transition rate which can be expressed as,
%   \begin{equation}
%      W(12;34) = \frac{d\sigma}{d\Omega} \delta( \mathbf{p}_{1} +\mathbf{p}_{2} -\mathbf{p}_{3} -\mathbf{p}_{4} ) \delta( \epsilon_{1} +\epsilon_{2} -\epsilon_{3} -\epsilon_{4} ).
%      \label{W}
%   \end{equation}
%   The term $\frac{d\sigma}{d\Omega}$ represents the differential cross-section of nucleon-nucleon scattering in the nuclear medium. The delta functions involving momenta $\mathbf{p}_{i}$ and single-particle energies $\epsilon_{i}$ (where $i$ = 1, 2, 3, and 4) reflect the conservation of momentum and energy during the scattering process. The single-particle energy $\epsilon_{i}$ is the sum of the single-particle kinetic energy and potential energy. The kinetic energy of a single particle is given by $\frac{pi^{2} }{2m}$, and the potential energy is $\int dr_{2}dp_{2}V_{12}   f_{1}(r_{2}  ,p_{2} )$.

In the numerical simulation of Eq.(\ref{BUU}), the parallel-ensemble method is employed, where N$_{\rm{tp}}$ systems are simulated simultaneously to obtain a smooth density distribution. In the literature, this approach is referred to as the BUU \cite{Bertsch1988}, Vlasov-Uehling-Uhlenbeck \cite{Aichelin1985, Kruse1985}, or Landau-Vlasov models \cite{Gregoire1987}.

The mean-field potential $U$ in Eq.(\ref{BUU}) includes the Coulomb and nuclear terms.
The Coulomb interaction between protons is described by simple point charges and calculated in each parallel-ensemble.
The specific form and parameters of the potential are detailed in the previous publication \cite{feng2023contributions}.
The nuclear potential parameters are derived from fitting the main features of the equation of state (EOS) for infinite nuclear matter, including saturation density, binding energy, incompressibility coefficient, symmetry energy and other characteristics.

At intermediate energies in heavy-ion collisions, nucleon-nucleon (NN) scattering is prevalent, and Pauli blocking in NN scattering is crucial for a proper quantum treatment, preventing a classical phase space distribution. 
However, when the incident energy is near the Coulomb barrier, the scenario undergoes a shift. 
In this energy regime, NN scattering becomes less frequent. 
Consequently, the NN collision term $I(f)$ in Eq.(\ref{BUU}) is implemented using the Phase-Space-Density Constraint (PSDC) method.
In the PSDC method, the phase space occupation probability is computed by accumulating the contributions of test particles in the BUU frame, as illustrated in the equation below.

\begin{equation}
  \overline{f}(\mathbf{r}_i,\mathbf{p}_i)=\frac{h^3}{2N_{tp}V_p}\sum_j^{V_p}\delta(\tau_i,\tau_j)\rho_{ij}
\end{equation}
where V$_{p}$ is the accumulating volume in the momentum, $\tau_{i}$ and $\tau_{j}$ are the isospin of the $i$th and $j$th test particles, $\delta$ function is for select the test particles with the same isospin, $\rho_{ij}$ is the density contribution of $j$th test particles in the position of the $i$th test particles.
Since the spin degree of freedom is not considered in the model, there is a number 2 in the denominator.

The PSDC method involves monitoring the phase-space occupation probability during time evolution. NN scatterings between neighboring nucleons are conducted to ensure that the phase-space occupation does not exceed 1 while minimizing the loss of reaction events. 
Momentum exchanges are selectively performed for nucleon pairs in which one nucleon has $\overline{f}(\mathbf{r}_i,\mathbf{p}_i)$ less than 0.3, and the other nucleon has $\overline{f}(\mathbf{r}_j,\mathbf{p}_j)$ greater than 1.05. 
The choice of 0.3 and 1.05 as empirical parameters is specific to this treatment.

In intermediate-energy heavy-ion collisions, the contribution of surface energy is negligible and can be disregarded. However, as the collision energy approaches the Coulomb barrier, the impact of surface energy becomes pronounced. To account for this, we introduced a surface energy constraint term into our model. 
The surface potential is
\begin{equation}
  U_{s} = g \nabla^{2} \rho,
\end{equation}
where $g$ is a parameter to adjust the strength of the surface potential. The value of $g$ chosen in this work is 5 MeV fm$^5$, which is the result obtained from the previous study \cite{feng2023contributions}. The inclusion of this constraint term ensures the preservation of the initial shape of the projectile and target nuclei for an extended duration in the BUU collision dynamics process, accounting for dissipative kinetic energy.

\begin{figure}
  \includegraphics[width=7.cm,angle=0]{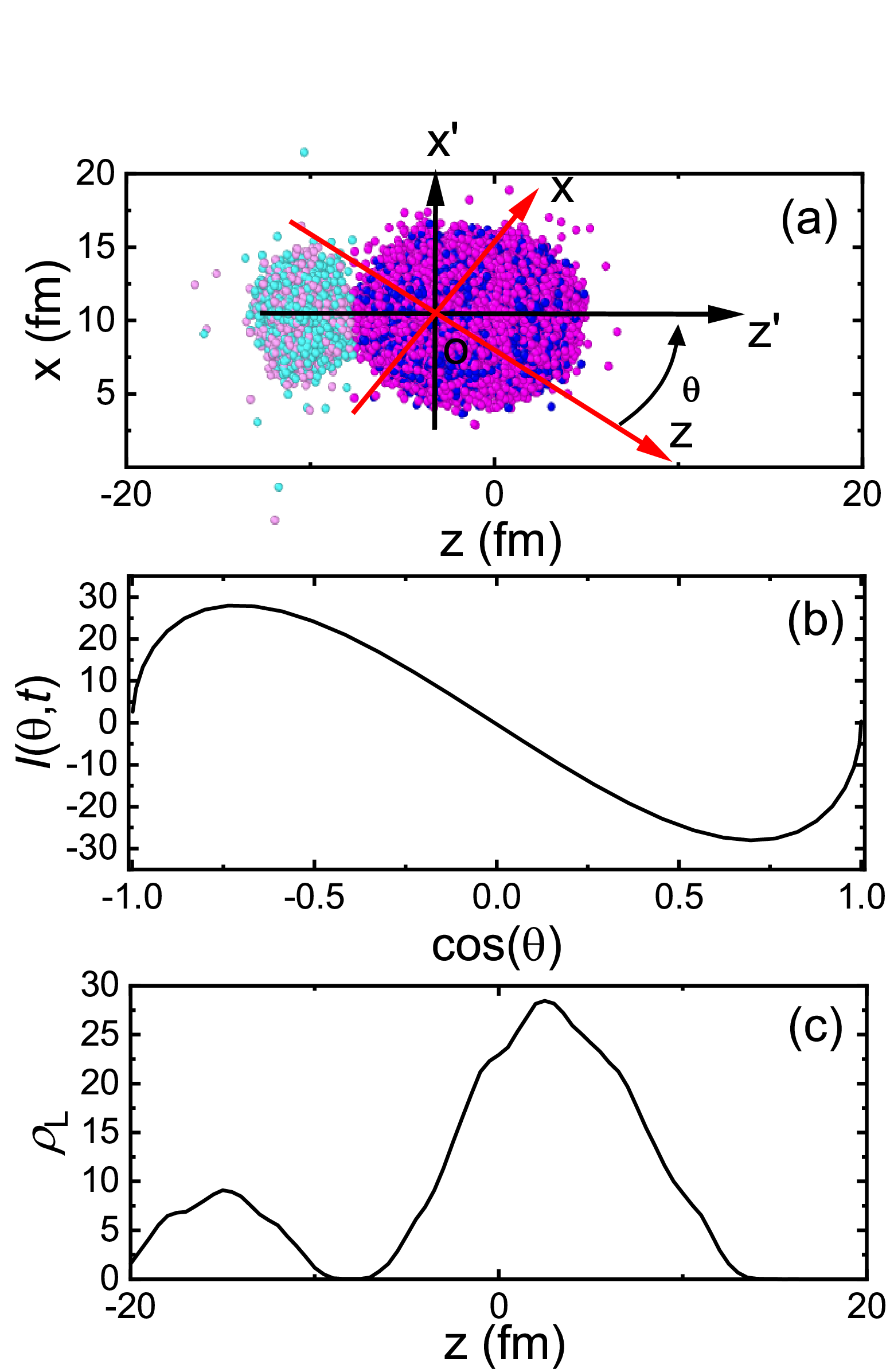}
  \captionsetup{justification=raggedright}
  \caption{\label{f1}(a)The setting of coordinate axes during the dynamical evolution of the dinuclear system involves. (b)The variation of the $\textit{I}(\theta,t)$ value as a function of the cosine value of the rotation angle of the coordinate axes. (c)The linear density distribution along the collision symmetry axis.}
\end{figure}

To investigate the dissipative dynamics in the QF process, it is essential to establish a window that distinguishes between the projectile and target nuclei at each time step. 
Initially, determining the symmetry axis of the dinuclear system is crucial, satisfying two conditions: passing through the center of mass of the dinuclear system and minimizing the rotational inertia of all test particles.
The $xz$ coordinate axis, with the center of mass of the dinuclear system as the origin, is established, as illustrated in Fig.\ref{f1}(a). At this point, the coordinates of all test particles are represented as ($z$, $x$). After rotation by an angle $\theta$, the coordinates of the test particles become ($z^{\prime}$, $x^{\prime}$). They satisfy the relationship:
$\begin{pmatrix}
  z'\\x'\end{pmatrix}$=$\begin{pmatrix}z\cos\theta+x\sin\theta\\x\cos\theta-z\sin\theta
\end{pmatrix}$
The $\textit{I}(\theta,t)$ values which is proportional to the rotational inertia is expressed in the following relationship:
\begin{equation}
  I(\theta,t)\propto\int x'z'\rho(\mathbf{r},t)d\mathbf{r}
\end{equation}
The relationship between $\textit{I}(\theta,t)$ and $\cos(\theta)$ is depicted in Fig.\ref{f1}(b). The $z$-axis that satisfies $\textit{I}(\theta,t) = 0$ serves as the symmetry axis of the dinuclear system, meeting the two aforementioned conditions.

Secondly, we calculate the linear density \textit{$\rho_{L}$} of the test particles along this symmetry axis, as illustrated in Fig.\ref{f1}(c). The \textit{$\rho_{L}$} distribution exhibits a double-peak shape, and the valley between the two peaks identifies the position of the window. A plane perpendicular to the symmetry axis at this position defines the window of the dinuclear system. By employing this window, researchers can investigate the temporal variations in mass, energy, and momentum of the projectile and target nuclei.

\begin{figure}
  \includegraphics[width=9.cm,angle=0]{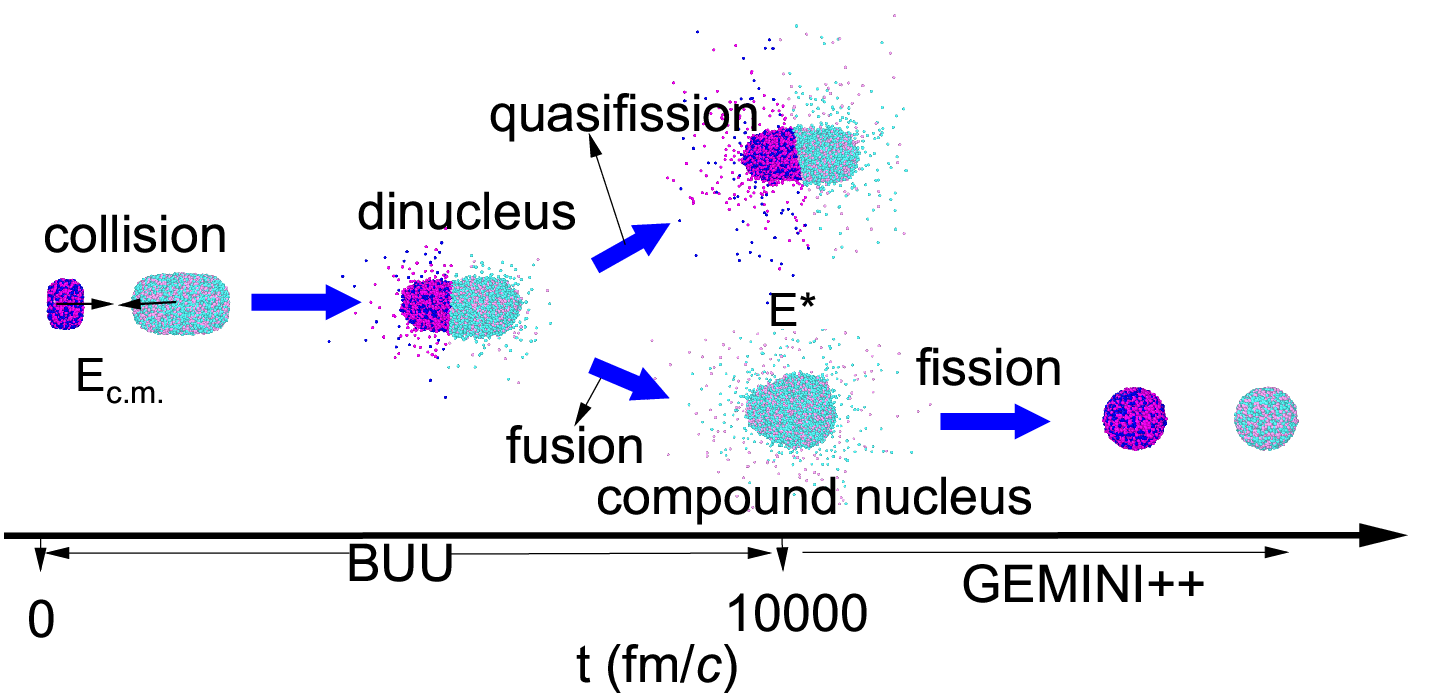}
  \captionsetup{justification=raggedright}
  \caption{\label{f2}The time range computed for the BUU model and GEMINI++ model}
\end{figure}
Figure \ref{f2} illustrates the models employed in this work at different time scales. The BUU model is utilized to simulate the dynamical process before the formation of the compound nucleus within a time range of 10000 fm/\textit{c}, encompassing QF reactions and fusion. Prior to 10000 fm/\textit{c}, the collision and capture of the projectile and target nuclei form a dinuclear system, which can subsequently undergo either QF or fusion reactions. The BUU model distinguishes between these two reactions by defining windows and provides the proportion of fusion and QF reactions through event-by-event simulations.
After 10000 fm/\textit{c}, the excited compound nuclei formed through fusion reactions undergo symmetric fission (referred to as FF), simulated using the GEMINI++ model \cite{charity2010systematic,mancusi2010unified}. By combining the BUU and GEMINI++ models, we can obtain the final-state fragments of the entire collision reaction, encompassing fragments from both QF and FF reactions.

\section{\label{method} Results and discussions }

\begin{figure*}
  \begin{minipage}[c]{0.3\linewidth}
    \centering%
    \includegraphics[width=1.\textwidth]{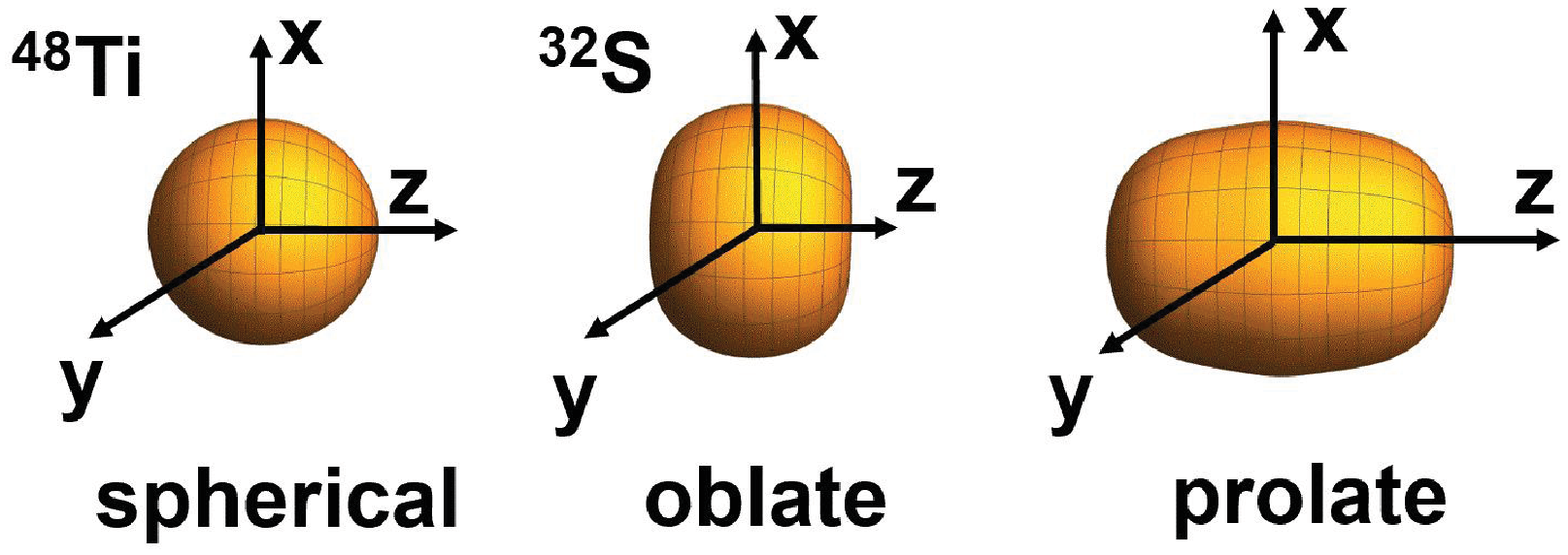}
    \label{f3}
    \end{minipage}%
    \hspace{1mm}%
    \begin{minipage}[c]{0.68\textwidth}
    \centering%
    \begin{tabular}{lccc}

    \hline
    \hline
    reaction system &$\beta_2^{\rm (P)}$ &$\beta_2^{\rm (T)}$ &shape of projectile nucleus and target nucleus  \\
    \hline
    $^{24}$Mg + $^{178}$Hf& 0.393 & 0.278  & prolate nuclei + prolate nuclei  \\
    \hline
    $^{34}$S + $^{168}$Er& - 0.235 & 0.297   & oblate nuclei + prolate nuclei  \\
    \hline
    $^{48}$Ti + $^{154}$Sm & 0.011 & 0.270 & spherical nuclei + prolate nuclei\\
    \hline
    \hline

    \end{tabular}
    \label{tab}
    \end{minipage}\\
    \begin{minipage}[t]{.3\textwidth}\centering%caption宏包
      \captionsetup{justification=raggedright}
    \captionof{figure}{The initial shape of nuclei for the three collision systems. The $^{34}$S nucleus is oblate. The $^{48}$Ti nucleus is spherical, while other nuclei are all prolate.}
    \end{minipage}\hspace{3mm}%
    \begin{minipage}[t]{.6\textwidth}\centering%
      \captionsetup{justification=raggedright}
    \captionof{table}{The $\beta_{2}$ values and their corresponding initial shape for the three collision systems. If the absolute value of the $\beta_{2}$ value is less than 0.1, the nucleus is approximately spherical. If the $\beta_{2}$ value is greater than 0.1, the nucleus is prolate. If the $\beta_{2}$ value is less than - 0.1, the nucleus is oblate.}
    \end{minipage}
  \end{figure*}

To investigate the effects of deformation and orientation on the competition between QF and FF, we selected collision systems with diverse deformations. This selection aimed to thoroughly examine the influence of initial nuclear deformations on the competition mechanism between QF and FF. Figure 3 displays the shapes of the selected collision nuclei in this study. The $^{48}$Ti nucleus is spherical, while the $^{34}$S nucleus is oblate, and the remaining nuclei are prolate.
Table I presents the combinations of the three collision systems and the $\beta_{2}$ values for each nucleus, along with the corresponding shape combinations for each selected collision system.

\begin{figure*}
  \includegraphics[width=14.cm,angle=0]{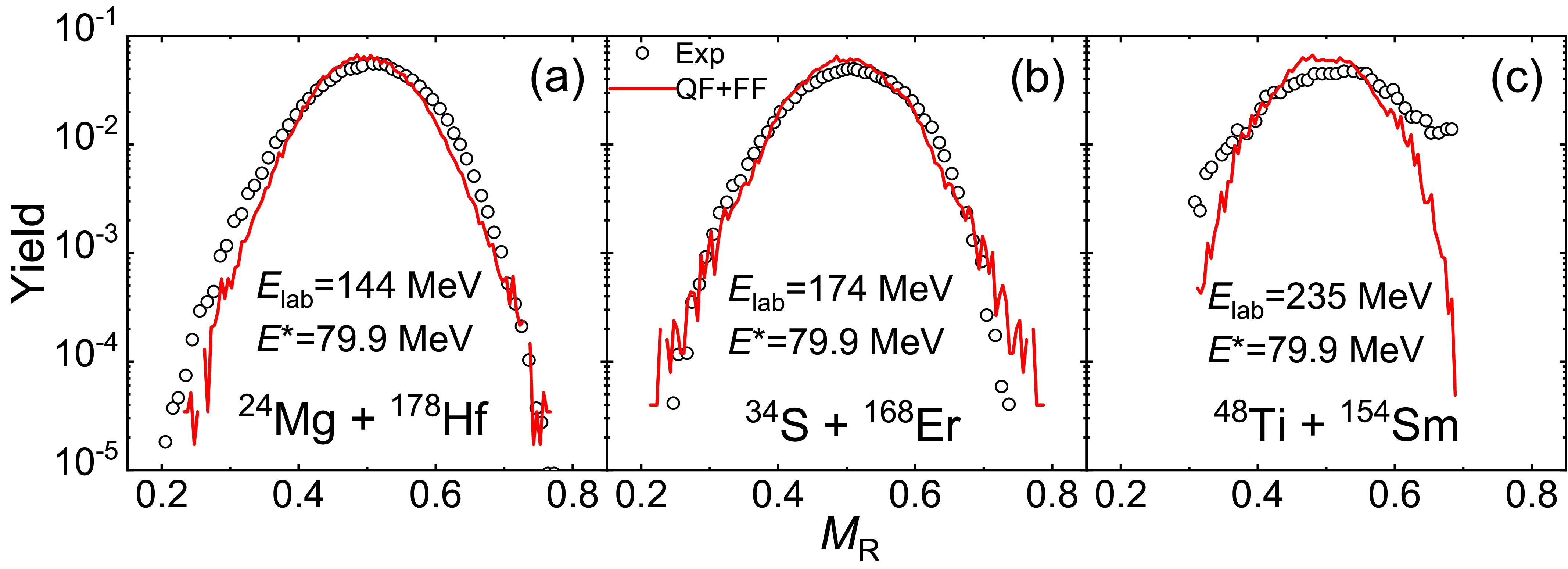}
  \captionsetup{justification=raggedright}
  \caption{\label{f4}Mass distributions of BUU + GEMINI++ calculations (QF + FF) of model (red line) and experimental data (circles) for (a) $^{24}$Mg+$^{178}$Hf, (b) $^{34}$S+$^{168}$Er, and (c) $^{48}$Ti + $^{154}$Sm at around 1.2 $V_B$. The incident energy (\textit{E}$_{lab}$) and the excitation energy of $^{202}$Po (The compound nucleus synthesized from the three systems) for each system are indicated. The experiment data and \textit{E}$_{lab}$ are from Ref. \cite{rafiei2008strong}, and the BUU model calculates the excitation energy. The $x$-axis corresponds to the mass ratio \textit{M}$_{R}$ = $\frac{m}{m1 + m2}$, where \textit{m} denotes the fragment mass number, and \textit{m$_{1}$} and \textit{m$_{2}$} represent the mass numbers of the projectile and target nuclei.}
\end{figure*}

In Figure \ref{f4}, a comparison is made between experimental data and model calculations of fragment mass distributions (MDs) for three distinct reaction systems. The MDs of QF are simulated using the BUU model, while the MDs of FF are simulated with the GEMINI++ model.
For the (a) $^{24}$Mg + $^{178}$Hf and (b) $^{34}$S + $^{168}$Er collisions, the model calculations align well with experimental data, demonstrating good agreement. However, some discrepancies are noted around \textit{M}${_R} \approx $ 0.7 and \textit{M}${_R} \approx $ 0.3. These deviations can be attributed to fluctuations in the BUU model simulations, and the presence of only a small fraction of fragments in this region.
Regarding the $^{48}$Ti + $^{154}$Sm collision, the model simulations match the experimental data for fragments in the $0.4 \leq \textit{M}{_R}\leqslant 0.6 $ range. Nevertheless, discrepancies are observed for fragments in the $0.3 \leq \textit{M}{_R}\leqslant 0.4 $ and $0.6 \leq \textit{M}_{R}\leqslant 0.7 $ ranges. This is attributed to the smaller mass asymmetry of the $^{48}$Ti + $^{154}$Sm system compared to the other two systems, leading to predominantly quasi-elastic and deep inelastic events that are not considered in our calculations.

Furthermore, the figure reveals that the three reactions form the same compound nucleus, $^{202}$Po, with identical excitation energy. However, the curves for both experimental and model data differ for different reaction systems, indicating variations in the MDs of QF for different systems. MDs of QF carry information about the entrance channel, influenced by factors such as the initial deformation of the projectile and target nuclei and the mass asymmetry of the projectile-target system, resulting in different MDs of QF for various reaction systems.

\begin{figure}
  \includegraphics[width=8.cm,angle=0 ]{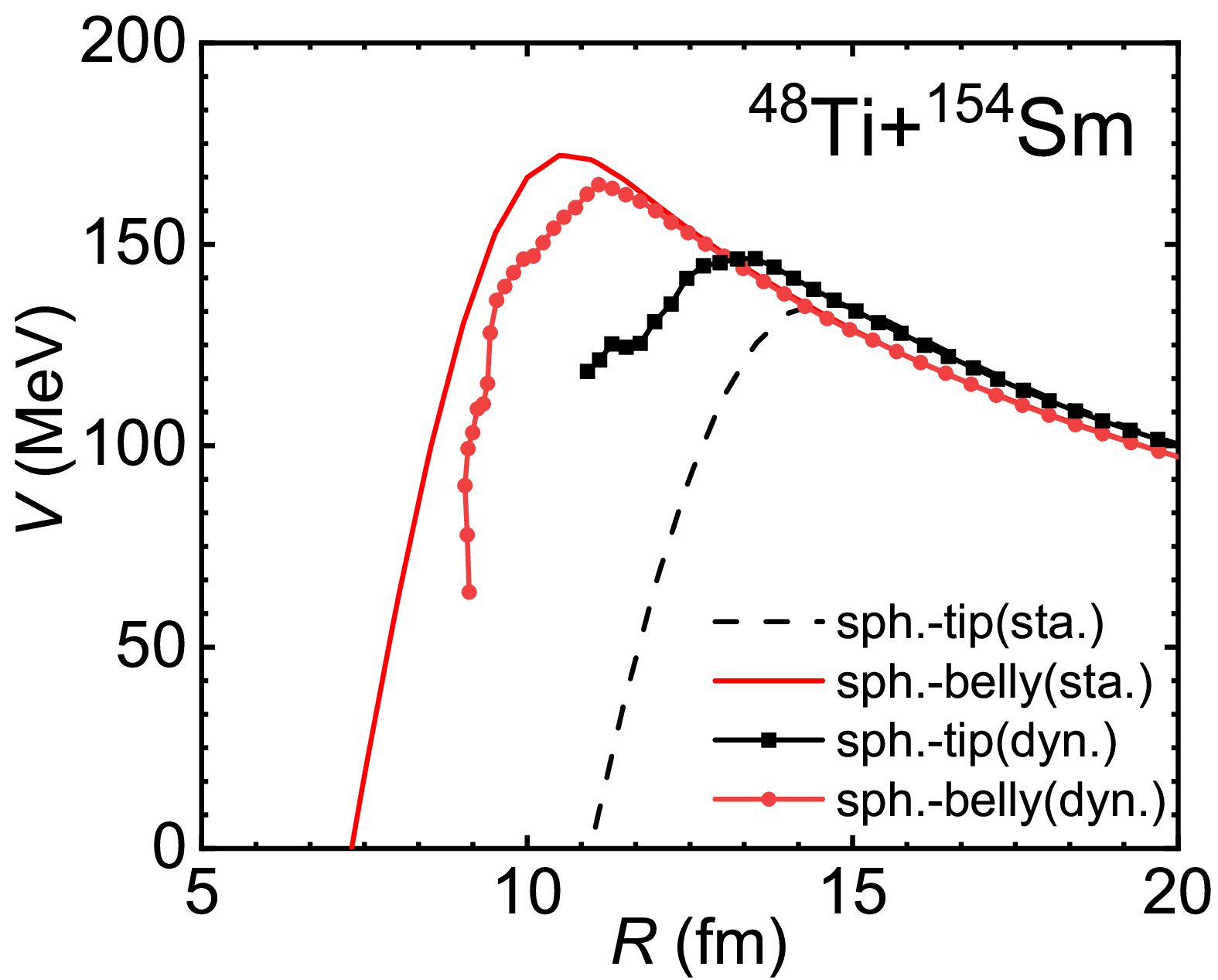}
  \captionsetup{justification=raggedright}
  \caption{
  \label{f5}
  (Color online) The dynamic and static nucleus-nucleus (NN) interaction curves for the $^{48}$Ti + $^{154}$Sm system, where the dynamic NN interactions calculation is the average result of simulating 30 events at an incident energy of \textit{E}$_{lab}$ = 235 MeV. The calculations in the figure are performed for central collisions. \textit{V} represents the NN interaction, and \textit{R} represents the center-of-mass distance between the two nuclei. The solid red lines represent the static NN interaction for the target nuclei in the belly orientation, while the solid red circle-dotted lines represent the dynamic one for the target in belly. The dashed black lines represent the static curve for the target in tip, and the solid red square-dotted lines represent the dynamic curve for the target in tip.}
\end{figure}

We understand that the orientation of deformed nuclei can impact the nucleus-nucleus interaction potential, subsequently affecting the capture probability and fusion cross-section of the projectile and target nuclei.
Fig. \ref{f5} depicts the static and dynamic interaction potential as a function of distance for the $^{48}$Ti + $^{154}$Sm system, considering the tip and belly orientations of the target nucleus. The interaction potential comprises the nucleus interaction potential $U_N$, the Coulomb interaction potential $U_C$, and the centrifugal potential. As these simulations are for central collisions, the centrifugal potential does not play a role.
It's observed that when the target nucleus is in the belly orientation, both the static and dynamic barriers are higher compared to the tip orientation. Moreover, the highest point of the interaction corresponds to a smaller \textit{R} compared to the tip orientation. This indicates that when the target nucleus is in the belly orientation, the collision results in a deeper contact between the projectile and target nuclei, leading to a larger $U_C$ and consequently a higher barrier.
Additionally, when the target nucleus is in the belly orientation, the static and dynamic interactions coincide at larger \textit{R}. However, at closer distances, the dynamic interaction is lower than the static interaction. This occurs because, at larger distances, the nucleon density of the projectile and target nuclei remains largely unchanged, resulting in the coincidence of the static and dynamic interactions. As the nuclei approach each other, dynamic changes in nuclear density lead to corresponding changes in dynamic interaction.

When the target nucleus is in the tip orientation, the dynamic and static interactions coincide at larger \textit{R}. However, at closer \textit{R}, the dynamic interaction surpasses the highest point of the static barrier, and the minimum distance at which the projectile and target nuclei can be in contact is smaller. The dynamic interaction for the tip orientation and the belly orientation of the target nucleus lies between the static interactions for the tip and belly orientations. This is attributed to the dynamic evolution of the BUU model, where nuclei cannot maintain their initial deformations and tend to evolve towards a spherical shape.
In conclusion, the initial orientation of deformed nuclei can influence the distribution of effective nucleus-nucleus interactions, aligning with conclusions from prior research \cite{hammerton2019entrance,li2005deformation}.

\begin{figure*}
  \includegraphics[width=16.cm,angle=0]{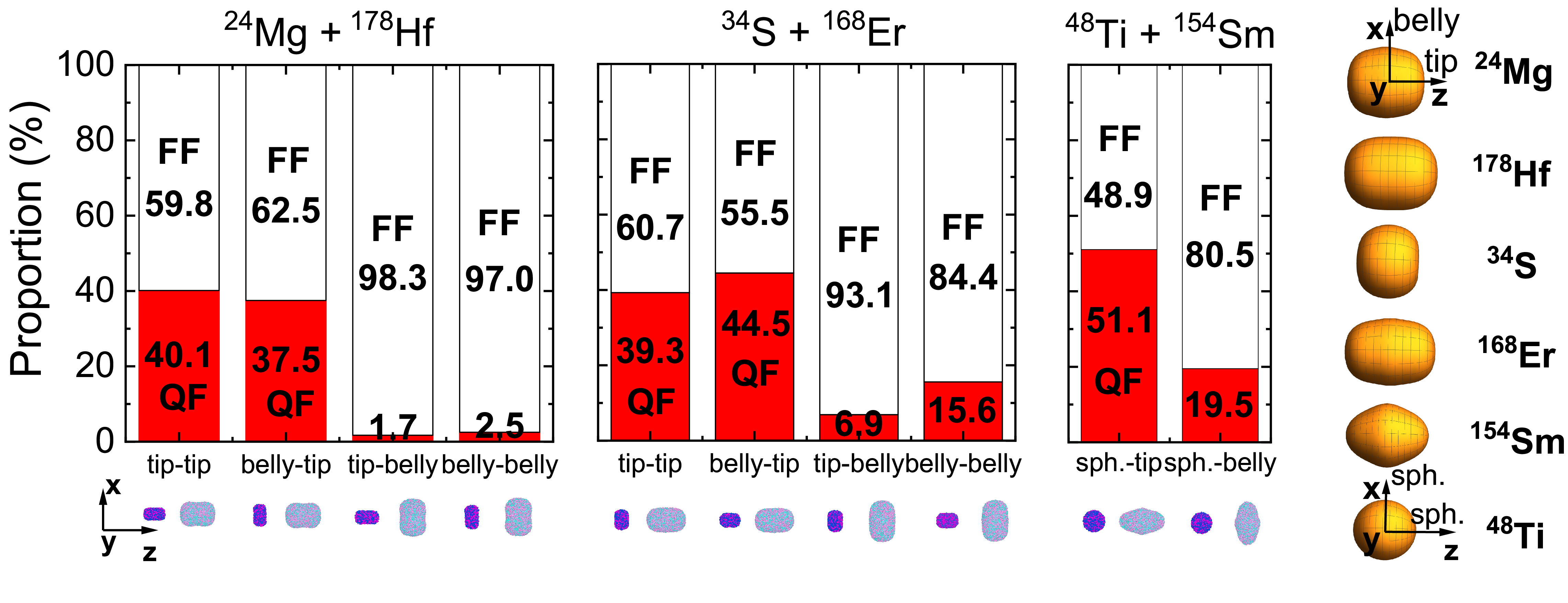}
  \captionsetup{justification=raggedright}
  \caption{\label{f6} The proportion of quasi-fission (QF) reaction and fusion-fission (FF) reaction for $^{24}$Mg + $^{178}$Hf, $^{34}$S + $^{168}$Er, and $^{48}$Ti + $^{154}$Sm  system in central collision at near Coulomb barrier energy, The initial deformation and orientation for projectile and target nuclei are indicated.
 }
\end{figure*}

Our calculations reveal that the initial orientation significantly influences the competition between QF and FF. 
In Figure \ref{f6}, we illustrate the proportions of QF and FF events among the total reaction events in $^{24}$Mg + $^{178}$Hf, $^{34}$S + $^{168}$Er, and $^{48}$Ti + $^{154}$Sm reactions for all possible combinations of orientations.
From the figure, it is evident that the competition between QF and FF is greatly affected by the orientation of the target nucleus. Specifically, in the three systems, the proportion of FF events is notably higher in the belly orientation of the target nuclei compared to the tip orientation.
Another noteworthy theoretical observation is that, as the charge product in the entrance channel (Z$_1$Z$_2$) increases, the overall proportion of QF events tends to increase. 
This phenomenon aligns with previous findings in the literature \cite{tsang1983energy}.

\begin{figure*}
  \includegraphics[width=15.cm,angle=0]{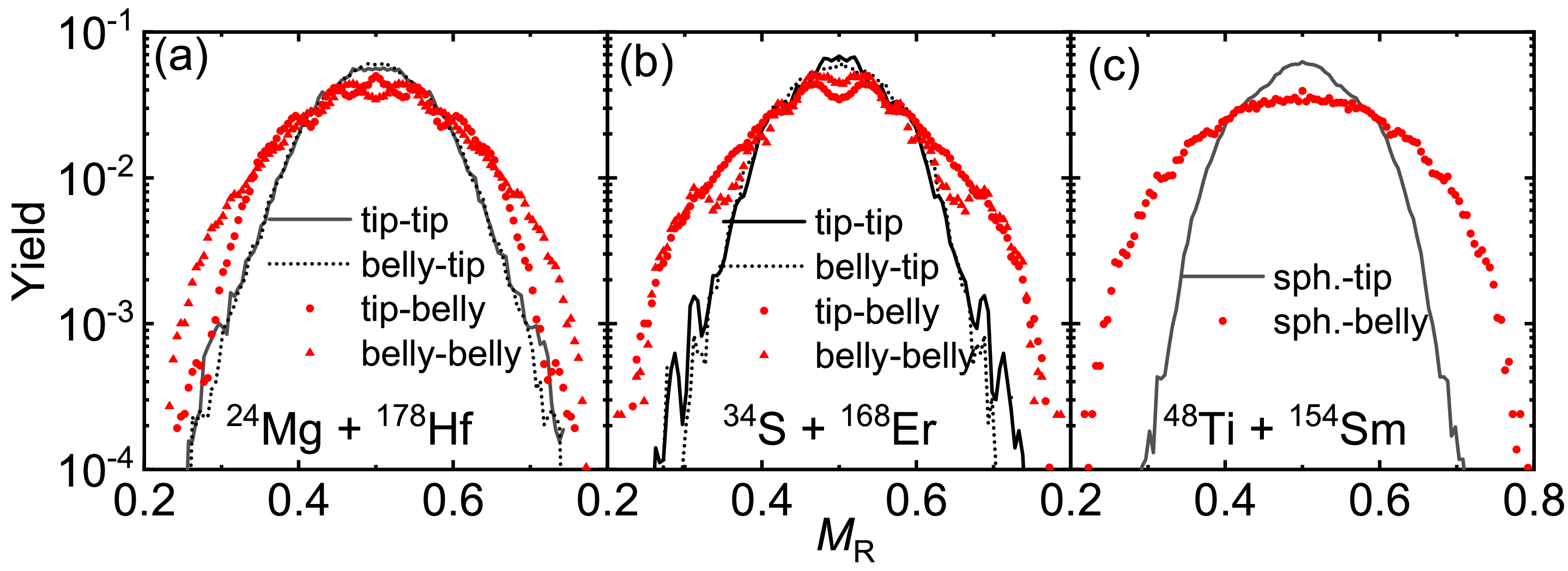}
  \captionsetup{justification=raggedright}
  \caption{
  \label{f7}
  (Color online)Mass distributions of quasi-fission (QF) fragments in (a) $^{24}$Mg+$^{178}$Hf, (b) $^{34}$S+$^{168}$Er, and (c) $^{48}$Ti + $^{154}$Sm in central collisions at near-Coulomb barrier with different orientations.
  }
\end{figure*}

On the other hand, collision configurations also influence the mass distributions of QF.
Figure \ref{f7} illustrates the mass distributions of QF reactions for (a) $^{24}$Mg + $^{178}$Hf, (b) $^{34}$S + $^{168}$Er, and (c) $^{48}$Ti + $^{154}$Sm under different orientations. The red dots represent the mass distributions of the target nucleus in the belly orientation, while the black line represents the mass distributions of the target nucleus in the tip orientation. It is evident that the width of the mass distribution is larger in the belly orientation of the target nucleus compared to the tip orientation.
In QF reactions, the width of the mass distributions can be attributed to the sticking time, referring to the survival time of the dinuclear system before separating into fragments \cite{rafiei2008strong, back2011experimental, shen1987fission}. Reactions with longer sticking times have a higher probability of forming fragments with a near-equilibrium mass ratio (M$_R$ = 0.5), resulting in narrower mass distributions. Conversely, shorter sticking times of the dinuclear system lead to wider mass distributions of QF fragments. Therefore, in the case of the target nucleus in the belly orientation, the sticking time of the dinuclear system is shorter compared to the tip orientation of the target nucleus. This factor contributes to influencing the competitive proportion between QF and FF.

\begin{figure}
  \includegraphics[width=9.cm,angle=0]{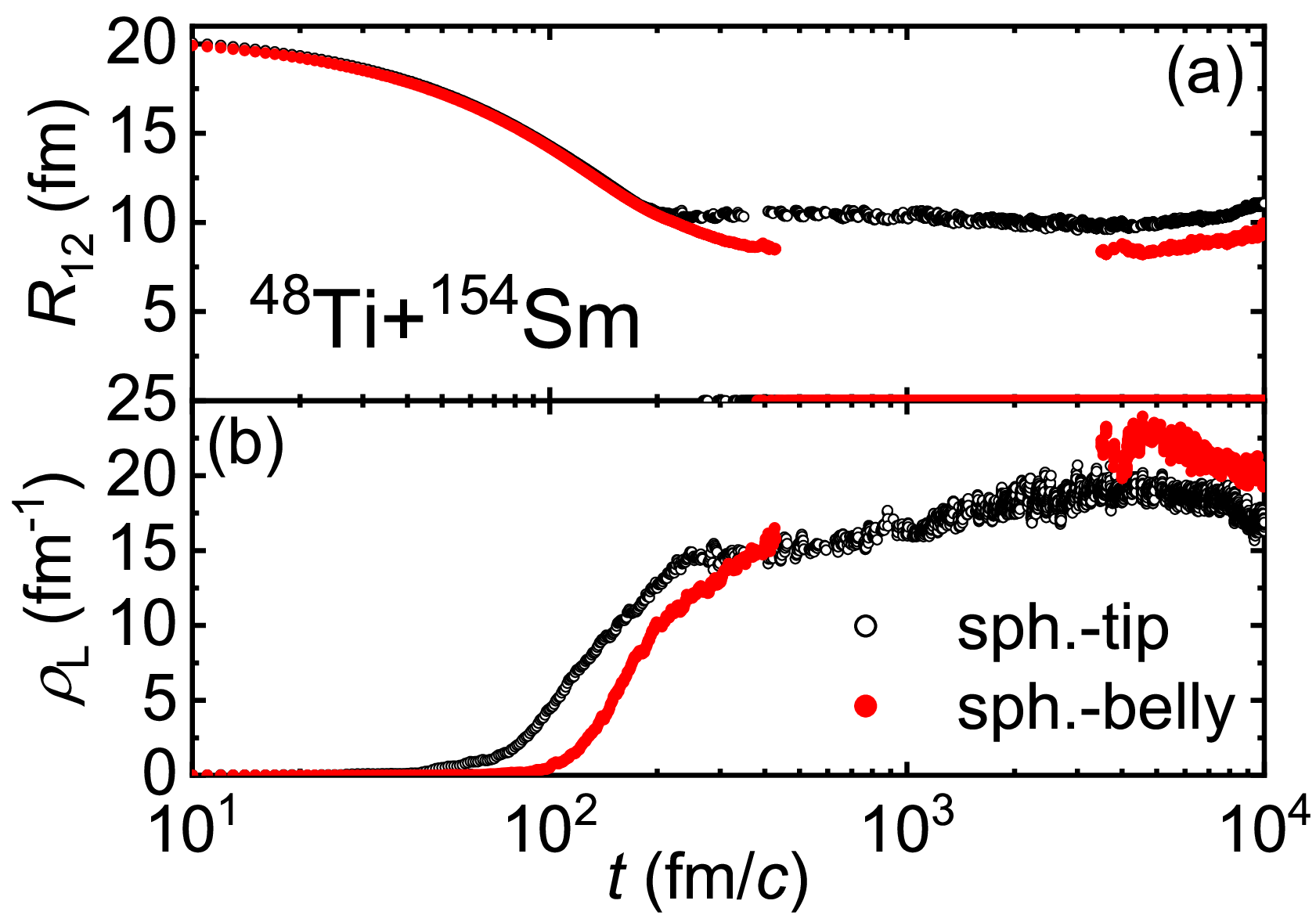}
  \captionsetup{justification=raggedright}
  \caption{ \label{f8}(Color online)The relationship between the center-of-mass distance \textit{R$_{12}$} of two nuclei and the linear density \textit{$\rho_{L}$} at the window for sph.-tip collisions (black hollow circles) and sph.-belly collisions (red solid circles) in the $^{48}$Ti + $^{154}$Sm system at \textit{E}$_{\rm lab}$ = 235 MeV as a function of reaction evolution time.}
\end{figure}

To investigate the microscopic dynamic mechanisms influencing the sticking time of the dinuclear system, we selected $^{48}$Ti + $^{154}$Sm, with different orientation combinations showing the maximum difference in width of mass distribution.
Figure \ref{f8} illustrates the evolution of the $^{48}$Ti + $^{154}$Sm system at an incident energy of \textit{E}$_{lab}$ = 235 MeV, comparing two orientations: target nucleus in the tip orientation (black hollow circles) and target nucleus in the belly orientation (red solid circles). The figure shows (a) the center-of-mass distance of the projectile and target nucleus \textit{R$_{12}$} and (b) the linear density \textit{$\rho_{L}$} at the window position as a function of reaction time.

Several distinct points are observable, the turning point of the curves, the extremum of the curves, and the duration of the curves at the extremum. Firstly, in Fig. \ref{f8} (b), the black line starts to rise at approximately 40 fm/\textit{c}, indicating the formation of the window around 40 fm/\textit{c} when the target nucleus is in the tip orientation. The red line starts to rise at around 100 fm/\textit{c}, signifying the formation of the window at approximately 100 fm/\textit{c} when the target nucleus is in the belly orientation. The sequential formation of the window is due to the elongated shape of the target nucleus, where in the tip orientation, the target nucleus comes into contact with the projectile nucleus earlier.

Secondly, the extremum of the curves in Fig. \ref{f8}(a) and (b) differ. In Fig. \ref{f8}(a), the minimum value that the red line can reach is around 7 fm, while the black line reaches a minimum of around 10 fm. This implies that collisions with the target nucleus in the belly orientation have a deeper contact with the projectile nucleus compared to collisions with the target nucleus in the tip orientation. The extremum of the curves in Fig. \ref{f8}(b) also confirms this point, showing that collisions with the target nucleus in the belly orientation exhibit a higher linear density at the window.

Thirdly, regarding the duration of the extremum, the black line reaches its extremum at 200 fm/\textit{c} and remains unchanged until 7000 fm/\textit{c}, while the red line reaches its extremum at around 400 fm/\textit{c} and remains unchanged until approximately 3500 fm/\textit{c}. Similarly, the duration shown in Fig. \ref{f8}(b) follows the same pattern. This indicates that the dinuclear system maintains a stable evolution for a longer time in sph.-tip compared to sph.-belly. This observation also aligns with the results in Fig. \ref{f6}, where the width of the MD in QF is related to the sticking time of the dinuclear system, with a wider mass distribution representing a shorter sticking time.

\begin{figure}
  \includegraphics[width=9.cm,angle=0]{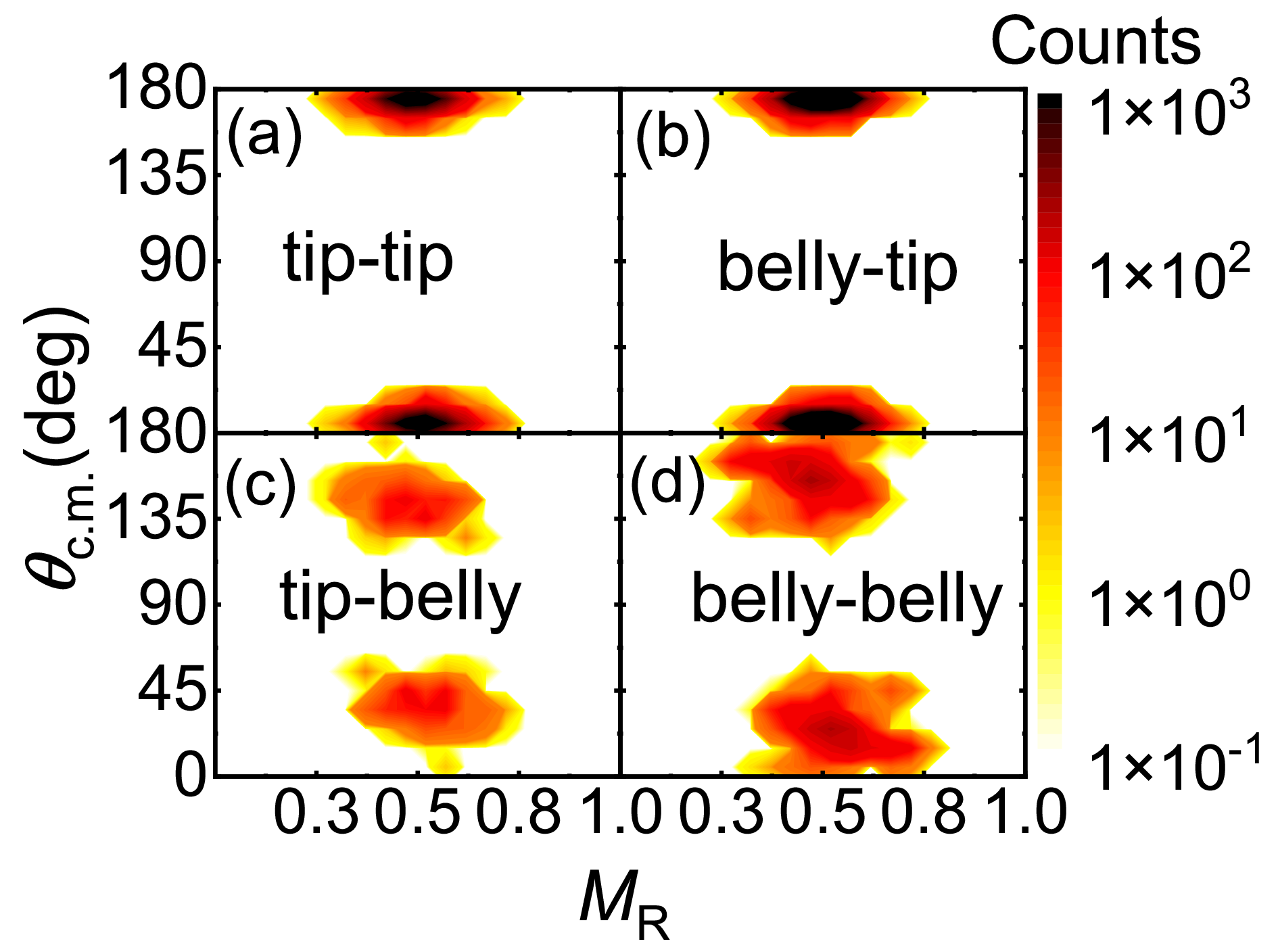}
  \captionsetup{justification=raggedright}
  \caption{\label{f9}
  (Color online)The mass-angle distributions of quasi-fission (QF) fragments for central collisions in different orientations of the $^{34}$S + $^{168}$Er system at \textit{E}$_{lab}$ = 174 MeV.}
\end{figure}

Figure \ref{f9} illustrates the mass-angle distributions for the $^{34}$S+$^{168}$Er system in central collisions at an incident energy of \textit{E}$_{\rm lab}$ = 174 MeV, considering four different collision orientations. The previously mentioned mass distribution are a subset of the mass-angle distribution, which not only provide information about the sticking time between the projectile and target nuclei through the fragmentation deflection angle but also offer insights into fluctuations between events.

Comparing (a) and (b) with (c) and (d) in Fig. \ref{f9}, three distinct points can be observed. Firstly, when the target nucleus is in the tip orientation, the reaction fragments are concentrated around M$_R$ = 0.5. On the other hand, when the target nucleus is in the belly orientation, the two reaction fragments are distributed on both sides of M$_R$ = 0.5. This implies that when the target nucleus is in the tip orientation, the dinuclear system in the reaction has a longer survival time compared to when the target nucleus is in the belly orientation. This indicates a more extensive exchange of nucleons between the projectile and target nuclei, resulting in a more symmetric MD of the two fragments after the reaction.

Secondly, when the target nucleus is in the tip orientation, QF fragments are concentrated at 0 degrees and 180 degrees, while for the belly orientation, QF fragments are concentrated at 30 degrees and 150 degrees.

Thirdly, the distribution of fragments is more concentrated when the target nucleus is in the tip orientation, whereas it is more dispersed when the target nucleus is in the belly orientation. This implies that events with the target nucleus in the tip orientation exhibit less fluctuation between events, while events with the target nucleus in the belly orientation have larger fluctuations. Therefore, the dynamical properties of each event and their uncertainties are greater when the target nucleus is in the belly orientation. The fluctuations between events are also a contributing factor to the dependence of the competition between QF and FF on the orientation of the target nucleus.

\begin{figure}
  \includegraphics[width=9.cm,angle=0]{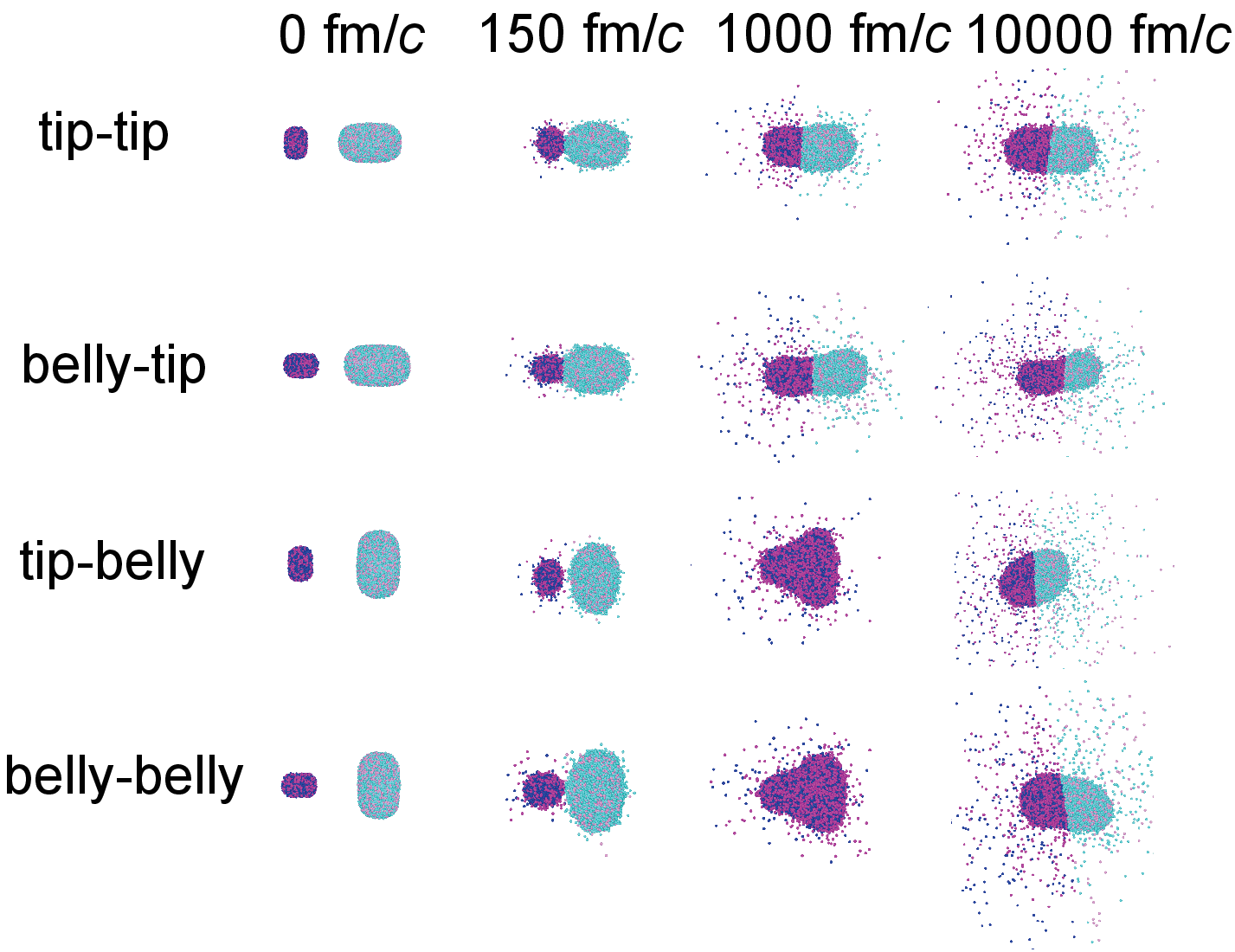}
  \captionsetup{justification=raggedright}
  \caption{
  \label{f10}
  (Color online)The phase space distribution of the $^{34}$S + $^{168}$Er system at different reaction times and orientations in central collisions with an incident energy of \textit{E}$_{lab}$ = 174 MeV.
  }
\end{figure}

Figure \ref{f10} shows the phase space distribution of the $^{34}$S + $^{168}$Er collision system at different orientations and moments in central collisions. From the figure, two pieces of information can be inferred related to the fluctuations between events.

Firstly, at a time of 1000 fm/c, the dinuclear system configuration for the tip-oriented target nucleus is elongated, while for the belly-oriented target nucleus, it is triangular. This implies that the triangular configuration of the dinuclear system is more unstable and will rapidly break apart into two fragments, resulting in larger uncertainties in the angles and masses of the emitted fragments.

Secondly, at a time of 10000 fm/c, the dinuclear system with the tip-oriented target nucleus exhibits minimal rotation relative to the incident direction, whereas the dinuclear system with the belly-oriented target nucleus undergoes a certain degree of rotation relative to the incident direction. This aligns with the previous observation that the triangular configuration of the dinuclear system is unstable. Therefore, this can explain the observed behavior in panels (c) and (d) of Figure \ref{f6}, where fragments obtained from collisions with the belly orientation of target nuclei exhibit a certain angular deflection and tend to be mass asymmetric. Additionally, there is a larger fluctuation between events in this type of collision.

\section{\label{summary}CONCLUSION}

In this study, we employed the Boltzmann-Uehling-Uhlenbeck (BUU) model to investigate the impact of orientation and deformation in heavy-ion collision reactions occurring around 1.2 times the Coulomb barrier (1.2 $V_B$). We selected three distinct shape combinations: $^{24}$Mg + $^{178}$Hf, $^{34}$S + $^{168}$Er, and $^{48}$Ti + $^{154}$Sm for our calculations.

Our investigation unveiled that different orientations of deformed nuclei have a significant influence on the nucleus-nucleus interaction potential. Specifically, the potential barrier is notably higher when the target nucleus is in the belly orientation. Furthermore, we established that the orientation of deformed target nuclei plays a pivotal role in determining the competition between QF and FF. A higher proportion of FF events is observed when the target nucleus is in the belly orientation compared to the tip orientation.
Additionally, our findings indicate that the orientation of the deformed target nucleus affects the width of the mass distributions of QF. Specifically, the mass distributions exhibit a wider spread when the target nucleus is in the belly orientation.

These observed phenomena are attributed to underlying micro-mechanisms, including the sticking time between the projectile and target nuclei and fluctuations between events in the reaction. Microscopic dynamical simulations using the BUU model revealed that the depth of contact between the projectile and target nuclei influences the survival time of the dinuclear system. Moreover, the configuration of the formed dinuclear system in the reaction is identified as a micro-mechanism impacting the fluctuations between events.

It should be emphasized that the BUU model can predict reaction dynamics in a nonempirical manner, as demonstrated in this study. Therefore, conducting further systematic BUU calculations for various projectile-
target combinations with different orientations, deformations, and entrance channel conditions has the potential
to provide effective reference settings for determining the initial conditions that influence fusion probabilities and
facilitate the experimental synthesis of new isotopes.

\section*{ACKNOWLEDGMENTS}

This work was supported by the National Natural Science Foundation of China under Grants Nos. 11875328 and 12075327, the Central Government Guidance Funds
for Local Scientific and Technological Development, China (No. Guike ZY22096024), the Key Laboratory
of Nuclear Data foundation(JCKY2022201C157).

\bibliography{reference}
\end{document}